# Magnetic field control of the near-field radiative heat transfer in three-body planar systems

Lei Qu[1,2], Edwin Moncada-Villa[3,*], Jie-Long Fang[1,2], Yong Zhang[1,2], Hong-Liang Yi[1,2,†]

[1]*School of Energy Science and Engineering, Harbin Institute of Technology, Harbin 150001, P. R. China*
[2]*Key Laboratory of Aerospace Thermophysics, Ministry of Industry and Information Technology, Harbin 150001, P. R. China*
[3]*Escuela de Física, Universidad Pedagógica y Tecnológica de Colombia, Avenida Central del Norte 39-115, Tunja, Colombia*

Recently, the application of an external magnetic field to actively control the near-field heat transfer has emerged as an appealing and promising technique. Existing studies have shown that an external static magnetic field tends to reduce the subwavelength radiative flux exchanged between two planar structures containing magneto-optical (MO) materials, but so far the near-field thermomagnetic effects in systems with more such structures at different temperatures have not been reported. Here, we are focused on examining how the presence of an external magnetic field modifies the radiative energy transfer in a many-body configuration consisting of three MO *n*-doped semiconductors slabs, separated by subwavelength vacuum gaps. To exactly calculate the radiative flux transferred in such an anisotropic planar system, a general Green-function-based approach is offered, which allows one to investigate the radiative heat transfer in arbitrary many-body systems with planar geometry. We demonstrate that, under specific choices of the geometrical and thermal parameters, the applied magnetic field is able to either reduce or enhance the near-field energy transfer in three-element MO planar systems, depending on the interplay between the damped evanescent fields of the zero-field surface waves and the propagating hyperbolic modes induced by magnetic fields. Our study broadens the understanding concerning to the use of external fields to actively control the heat transfer in subwavelength regimes, and may be leveraged for potential applications in the realm of nanoscale thermal management.

## I. INTRODUCTION

Two objects at different temperatures and separated by a vacuum gap exchange heat via electromagnetic waves. This radiative heat transfer (RHT) cannot exceed the limit set by the Stefan-Boltzmann law if the separation between objects is greater than the thermal wavelength $\lambda_T = \hbar c / k_B T$. If objects are brought in closest proximity, at a distance below such characteristic length, an additional contribution, known as the near-field radiative heat transfer (NFRHT), arising from the evanescent field of the surface waves, can overcome by several orders of magnitude the far-field heat transfer by propagating waves [1–6]. This new contribution to the RHT has led to the research of functional devices for modern energy technologies, such as thermophotovoltaics [7–10], thermal lithography [11], heat-assisted magnetic recording [12], and scanning thermal microscopy [13].

The past decades have seen a growth of interest in further enhancing the NFRHT by the engineering of photonic structures or the implementation of materials whose optical properties are optimal for the NFRHT [14–19]. In parallel, a long-standing problem is the quest for the active control of extraordinary near-field electromagnetic energy. For this, many innovative strategies have been proposed [20–27]. One of the most appealing and promising scenarios is utilizing an external static magnetic field to actively tune the near-field thermal exchange between *n*-doped semiconductors like InSb or Si [6]. Depending on their doping level, the optical properties of these materials are very sensitive to the action of an external magnetic field, leading to strong magneto-optical (MO) activities in the infrared region of the electromagnetic spectra (relevant for thermal radiation). This strategy for controlling the NFRHT was first proposed by Moncada-Villa *et al*. in Ref. [24]. It was shown theoretically that the RHT in the near-field regime, between two identical planar semi-infinite slabs made of MO semiconductors, is always reduced by the application of a static magnetic field. Such a decrease is the consequence of the replacement of surface polariton modes (evanescent inside the slabs as well as in vacuum), which dominate the zero-field heat exchange, by hyperbolic modes (propagating inside the slabs and evanescent in vacuum) which have a lower tunneling probability. Later, magnetic field effects were also reported for the NFRHT in another symmetric systems as two identical MO nanoparticles [28,29] or magnetophotonic crystals [30,31], for which an optimal choice of the geometrical parameters (e.g., film thickness) leads to a reduction of the NFRHT being greater than the corresponding one to two semi-infinite InSb slabs. In contrast, only a few studies have reported the possibility of the magnetic field enhancement of the NFRHT in non-symmetric planar configurations, as a semi-infinite slab made of InSb separated by a vacuum gap of a different material like gold [32] or graphene [33]. The increase of the subwavelength radiative flux in such configurations is because the field-induced hyperbolic modes do not spoil the zero-field surface cavity modes due to the mismatch of resonance frequencies of dissimilar materials.

---

* edwin.moncada@uptc.edu.co
† yihongliang@hit.edu.cn

The above-mentioned works on the magnetic field dependence of the NFRHT have been confined to two-body systems. However, many-body problems are indispensable to construct frameworks for many branches of physics, including condensed matter physics, atomic physics, and so on [34,35]. In an experimental setup, the RHT may takes place in systems containing three or more bodies, where the novel physical and transfer phenomenon with no analogues in two-body systems will arise [36–45]. Particularly in systems consisting of more than two MO nanoparticles, researchers have predicted some interesting near-field thermomagnetic phenomena, such as the thermal Hall effect [46], the persistent heat current in thermal equilibrium [47], and the giant thermal magnetoresistance for the heat flux [48], which are not exhibited by two-body configurations. In addition, the features of the NFRHT between many planar bodies, comprised by isotropic dissimilar materials, have provided the possibility of the tailoring of functional devices for thermal management following the operating principle of electronic devices [49–54]. One example of this kind of devices is the near-field thermal transistor, allowing the amplification of the heat flux from a hot (source) slab to a cold (drain) one by manipulating the temperature of a third body placed between them, as well as the geometrical parameters [49,52].

Despite numerous efforts dedicated to the many-body radiative heat transfer, the magnetic field effects upon the NFRHT in systems with three or more MO planar bodies have not been reported so far. Notably, when exposed to external magnetic fields, MO materials such as doped semiconductors is optically anisotropic and its permittivity tensor has complex off-diagonal components, which renders many-body planar systems comprising such materials inherently different from previously reported ones that only consist of isotropic or uniaxial media [37–39,49–54].

In this work, we theoretically investigate how the presence of an external static magnetic field influences the near-field heat transfer in a many-body system, consisting of three noncontact parallel slabs of MO *n*-doped InSb [see Fig.1]. We provide a general Green-function-based approach for the investigation of the RHT in arbitrary many-body planar configurations, which allows us to exactly calculate the radiative flux transferred in anisotropic three-body systems we consider. It is shown that, under certain choices of temperatures and geometrical parameters, the inclusion of the intermediate slab may lead either to a reduction or an increment of the NFRHT by applying a magnetic field. Our finding could be of importance for developing the magnetic field control of electromagnetic energy transfer in complex systems of macroscopic bodies.

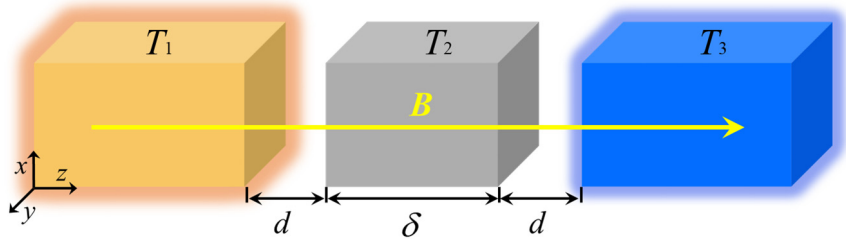

**Fig. 1.** Schematic representation of three parallel slabs made of *n*-doped InSb in the presence of an external static magnetic field parallel to the energy transport direction. An intermediate slab with thickness $\delta$ is placed in between the outermost slabs, and separated by vacuum gaps of size *d*.

The remainder of this paper is organized as follows. In Sec. II, we give a description of the system under consideration. In Sec. III, we introduce the theoretical approach for the calculation of the RHT in arbitrary many-body planar systems. Section IV is devoted to the analysis of the mechanisms underlying the magnetic field reduction/enhancement of the NFRHT in the MO three-body system near the thermal equilibrium, and, in Sec. V we briefly discuss the magnetic field effects upon the radiative energy flux in steady state. Our main results are summarized in Sec. VI.

**II. PHYSICAL SYSTEM UNDER STUDY**

The goal of this paper is the calculation of the NFRHT in a system out of thermal equilibrium, consisting of three noncontact planar slabs made of MO *n*-doped InSb, in the presence of an external static magnetic field **B** [see Fig. 1]. Outermost slabs 1 and 3, assumed as semi-infinite, are held at temperatures $T_1$ and $T_3$ ($T_1 > T_3$), respectively, whereas the slab placed between them has a temperature $T_2$ and a thickness $\delta$. This slab is separated by vacuum gaps of size *d* from the outermost slabs. From now on, we assume that the orientation of the applied magnetic field is perpendicular to the surface of the slabs (i.e., $\mathbf{B} = B\hat{z}$). It should be mentioned that the rotation the field orientation does not radically change the radiation property of planar InSb [32]. In the presence of an external field oriented as $\mathbf{B} = B\hat{z}$, InSb exhibits an optical anisotropy described by the following dielectric permittivity tensor [55]

$$\hat{\varepsilon}_{\text{InSb}} = \begin{bmatrix} \varepsilon_1 & -i\varepsilon_2 & 0 \\ i\varepsilon_2 & \varepsilon_1 & 0 \\ 0 & 0 & \varepsilon_3 \end{bmatrix}, \tag{1}$$

where

$$\begin{aligned} \varepsilon_1(B) &= \varepsilon_\infty \left( 1 + \frac{\omega_L^2 - \omega_T^2}{\omega_T^2 - \omega^2 - i\Gamma\omega} + \frac{\omega_p^2(\omega + i\gamma)}{\omega\left[\omega_c^2 - (\omega + i\gamma)^2\right]} \right), \\ \varepsilon_2(B) &= \frac{\varepsilon_\infty \omega_p^2 \omega_c}{\omega\left[(\omega + i\gamma)^2 - \omega_c^2\right]}, \\ \varepsilon_3 &= \varepsilon_\infty \left( 1 + \frac{\omega_L^2 - \omega_T^2}{\omega_T^2 - \omega^2 - i\Gamma\omega} + \frac{\omega_p^2}{\omega(\omega + i\gamma)} \right), \end{aligned} \tag{2}$$

Here, $\varepsilon_\infty = 15.7$ is the high frequency permittivity, $\omega_L = 3.62 \times 10^{13}$ rad s$^{-1}$ ($\omega_T = 3.39 \times 10^{13}$ rad s$^{-1}$) is the longitudinal (transversal) phonon frequency, $\omega_p = 3.14 \times 10^{13}$ rad s$^{-1}$ is the plasma frequency, and, $\Gamma = 5.65 \times 10^{11}$ rad s$^{-1}$ ($\gamma = 3.39 \times 10^{12}$ rad s$^{-1}$) is the phonon (free carrier) damping constants. The effect of the magnetic field is expressed by cyclotron frequency $\omega_c = eB/m^*$, where $m^* = 1.99 \times 10^{-32}$ kg is the effective mass corresponding to a doping level of $1.07 \times 10^{17}$ cm$^{-3}$.

In the absence of magnetic fields, $\varepsilon_1 = \varepsilon_3$ and $\varepsilon_2 = 0$, and thus, the InSb is optically isotropic. In this case, the interface between the InSb and the vacuum, can support either surface plasmon polaritons (SPPs) at frequencies below the surface plasmon frequency $\omega_{spp} = \omega_p/\sqrt{2}$, or surface phonon polaritons (SPhPs) in the restralhen band $\omega_L < \omega < \omega_T$ [55]. These modes are characterized by an evanescent field inside both InSb and vacuum, and by having a high component of the wave vector parallel to the interfaces, favoring the strong thermal exchange. When the magnetic field is turned on, magneto-optical effects are induced via the off-diagonal element, $\varepsilon_2$. More importantly, for some frequency regions a new kind of electromagnetic modes emerge. These modes are referred as hyperbolic modes (HM) and classified as type I hyperbolic modes (HMI) if $\varepsilon_1 > 0$ and $\varepsilon_3 < 0$, and type II (HMII) if $\varepsilon_1 < 0$ and $\varepsilon_3 > 0$, as has been discussed in Ref. [23,30]. Hyperbolic modes are propagating inside MO materials and evanescent in vacuum, and have a component of the wave vector parallel to the interfaces smaller than the one corresponding to the surface waves. This leads to a reduction of the NFRHT by the action of a magnetic field upon two semi-infinite InSb plates, due to the replacement of SPP and SPhP modes by hyperbolic modes [23].

We focus on the analysis of the net heat flux, $\varphi_3$, received by the cold body (i.e., slab 3), which is rigorously given by the following Landauer-like expression [47,48]

$$\varphi_3 = \int_0^\infty \frac{d\omega}{2\pi} \int \frac{d\mathbf{k}}{(2\pi)^2} \sum_{j=1,2} \left[ \Theta_j(\omega, T_j) \mathcal{T}_{j \to 3}(\omega, \mathbf{k}) - \Theta_3(\omega, T_3) \mathcal{T}_{3 \to j}(\omega, \mathbf{k}) \right], \tag{3}$$

where index $j = 1, 2$, $\Theta_j(\omega, T_j) = \hbar\omega/[\exp(\hbar\omega/k_B T_j) - 1]$ is the mean energy of photons with frequency $\omega$, $\mathbf{k} = (k_x, k_y)$ is the component of wave vector parallel to the interfaces, and $\mathcal{T}_{j \to 3}$ ($\mathcal{T}_{3 \to j}$) denotes the transmission probability for thermal radiation coming from slab $j$ to slab 3 (slab 3 to slab $j$), whose calculation is described in next section.

## III. THEORETICAL APPROACH

As mentioned in the Introduction, existing works on the NFRHT in many-body planar systems only consists of isotropic or uniaxial materials (with the optical axis parallel to the energy transfer direction). In such systems, the transmission probability can be calculated by using the analytical formula derived from the scattering approach [37–39]. This formula, however, is not suitable for the exact calculation of the photon transmission probability in many-body systems contains optically anisotropic materials of which permittivity tensor has off-diagonal components (e.g., MO materials), because it does not include the contributions from polarization conversion (i.e., TM→TE and TE→TM).

In fact, since thermal radiation originates from random fluctuating current sources, directly implementing the volume integration over all these sources is a very effective method to investigate the NFRHT [56,57]. In such an approach, the thermal radiation fields are linked to the fluctuating current sources via the Green's function. For layered structures, as shown by Francoeur. et al [56], this method allows each layer to have different temperatures (in this case layered systems can be regarded as many-body systems), and can compute the transmission probability between any two layers (bodies) in planar configurations. Besides, it allows one to trace the radiative heat flux at any position in systems, which is unattainable for the scattering approach [37–39]. Nevertheless, the algorithm provided in Ref. [56] is only applicable to isotropic materials. In the following, we will generalize this method to the case where layered systems may comprise any kind of optically anisotropic materials.

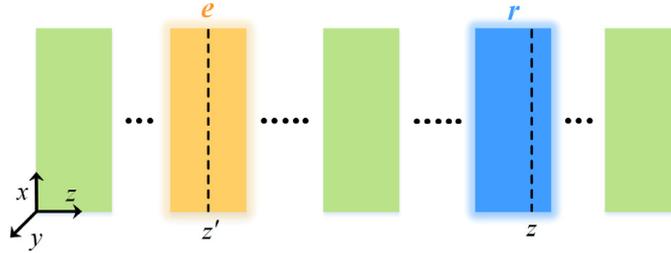

**Fig. 2.** Schematics of the system consisting of any number of layers (bodies). These layers may have different thicknesses and temperatures. Thermal emission of fluctuating currents contained in the $z'$ plane in layer $e$, is absorbed by the $z$ plane in the receiving layer, $r$.

Let us consider a system consisting of multiple layers stacked in the $z$ direction and infinite along the $x$ and $y$ directions, as sketched in Fig. 2. These layers may have different temperatures, and be made of arbitrary materials with the permittivity tensor $\bar{\varepsilon}$. As an example, we focus on analyzing the situation where the energy flux is transferred from the emitting layer $e$ to the receiving layer $r$. To determine the transmission probability between them, it is necessary to know the radiative flux coming from layer $e$ to the two boundaries of layer $r$. For this, we are going to calculate the flux at arbitrary position $z$ within layer $r$, generated from the fluctuating current sources within layer $e$, which is given by the $z$-component Poynting vector

$$\varphi_{e \to z}(\mathbf{r},t) = S(\mathbf{r},z,t) = \hat{\mathbf{z}} \cdot \langle \mathbf{E}(\mathbf{r},z,t) \times \mathbf{H}(\mathbf{r},z,t) \rangle, \tag{4}$$

where $\langle \cdots \rangle$ denotes the ensemble average, $\mathbf{E}(\mathbf{r}, z, t)$ and $\mathbf{H}(\mathbf{r}, z, t)$ are the electric and magnetic fields, and $\mathbf{r}$ and $t$, respectively, denotes the in-plane coordinate and time. Using the Fourier transform in time and space, defined as $f(t) = \text{Re} \int_0^\infty d\omega f(\omega) e^{-i\omega t}$ and $f(\mathbf{r}) = \int \frac{d\mathbf{k}}{(2\pi)^2} f(\mathbf{k}) e^{i\mathbf{k} \cdot \mathbf{r}}$, the $z$-direction Poynting vector can be rewritten as

$$S = \frac{1}{2} \text{Re} \int_0^\infty d\omega \int \frac{d\mathbf{k}}{(2\pi)^4} \hat{\mathbf{z}} \cdot \langle \mathbf{E}(\mathbf{k}, z, \omega) \times \mathbf{H}^*(\mathbf{k}, z, \omega) \rangle. \tag{5}$$

To facilitate the understanding and calculation, we recast Eq. (5) into the following form:

$$S = \frac{1}{2} \text{Re} \int_0^\infty d\omega \int \frac{d\mathbf{k}}{(2\pi)^4} \text{Tr}\left[ \hat{\Gamma} \langle \mathbf{E}\mathbf{H}^* \rangle \right]. \tag{6}$$

Here, matrix $\hat{\Gamma} = \begin{bmatrix} 0 & -1 & 0 \\ 1 & 0 & 0 \\ 0 & 0 & 0 \end{bmatrix}$, and $\langle \mathbf{E}\mathbf{H}^* \rangle = \langle \mathbf{E}_\alpha \mathbf{H}_\beta^* \rangle$, where $\alpha$ and $\beta$ represent the Cartesian components. We note that the arguments of $\mathbf{k}$, $z$, and $\omega$ have been suppressed.

For each $\mathbf{k}$ and $\omega$, we express the thermally generated electric and magnetic fields at position $z$ by considering all contributions arising from the fluctuating current sources, $\mathbf{J}(\mathbf{k}, \omega, z')$, at the $z'$ plane within layer $e$ as

$$\mathbf{E} = \int_{z'} \hat{G}_E \mathbf{J}(\mathbf{k}, \omega, z'), \tag{7}$$

$$\mathbf{H} = \int_{z'} \hat{G}_H \mathbf{J}(\mathbf{k}, \omega, z'), \tag{8}$$

where $\hat{G}_E$ and $\hat{G}_H$ are the electric and magnetic Green's function, which connects the emitting sources in a plane located at position $z'$ in the emitting layer, with the absorbing plane at the position $z$ in the receiving layer $r$. It is worthy to mention that, for the convenience of our calculation, this expression defined here is somewhat different from that in Refs. [56,57]. By using the scattering matrix formalism [58], we can exactly calculate these Green's functions. For details of the calculation, we refer the reader to Supplemental Material [59].

According to Eqs. (6)–(8), it becomes clear that solving the heat flux requires the calculation of ensemble average of the spatial correlation function of $\mathbf{J}$ within layer $e$, which can be given by the fluctuation-dissipation theorem [2]

$$\langle \mathbf{J}(\mathbf{k}, \omega, z) \mathbf{J}^\dagger(\mathbf{k}', \omega, z') \rangle = (2\pi)^2 \frac{4}{\pi} \omega \varepsilon_0 \Theta(\omega, T_e) \frac{\hat{\varepsilon}_e - \hat{\varepsilon}_e^\dagger}{2i} \delta(\mathbf{k} - \mathbf{k}') \delta(z - z'), \tag{9}$$

where $T_e$ and $\hat{\varepsilon}_e$ are, respectively, the temperature and the permittivity tensor of the emitting layer $e$.

Plugging Eqs. (7)–(9) into Eq. (6), we generate the following expression for the radiative heat flux at position $z$ within the receiving layer $r$:

$$S = \int_0^\infty \frac{d\omega}{2\pi} \int \frac{d\mathbf{k}}{(2\pi)^2} 4\omega \varepsilon_0 \Theta(\omega, T) \int_{z'} \text{Re} \text{Tr}\left[ \hat{\Gamma} \hat{G}_E \frac{\hat{\varepsilon} - \hat{\varepsilon}^\dagger}{2i} \hat{G}_H^\dagger \right]. \tag{10}$$

Accordingly, the photon transmission probability can be written as

$$\mathcal{T}(\omega, \mathbf{k}) = 4\omega \varepsilon_0 \int_{z'} \text{Re} \text{Tr}\left[ \hat{\Gamma} \hat{G}_E \frac{\hat{\varepsilon} - \hat{\varepsilon}^\dagger}{2i} \hat{G}_H^\dagger \right]. \tag{11}$$

We note that Eq. (11) allows one to calculate the transmission probability from the emitting layer to any position *z* within the receiving layer, thereby guaranteeing that the transmission probability between these two layers can be solved.

Let us finish this section by saying that the numerical simulations for the MO three-body system we consider have shown that $\mathcal{T}_{j\to 3}(\omega,\mathbf{k}) = \mathcal{T}_{3\to j}(\omega,\mathbf{k})$, with *j* =1, 2. This reciprocity of the heat flux received/emitted by slab 3 is consistent with the results reported in Ref. [48] for a linear chain of InSb nanoparticles. With this in mind, Eq. (3) can be simplified as follows:

$$\varphi_3 = \int_0^\infty \frac{d\omega}{2\pi} \int \frac{d\mathbf{k}}{(2\pi)^2} \sum_{j=1,2} \left[\Theta_j(\omega,T_j) - \Theta_3(\omega,T_3)\right] \mathcal{T}_{j\to 3}(\omega,\mathbf{k}). \qquad (12)$$

We employ this relation to analyze the magnetic field dependence of the RHT in the system of Fig. 1, for two different choices of temperatures. The first one, analyzed in next section, corresponds to the thermal equilibrium between slabs 2 and 3, i.e., the intermediate slab 2 purely behaving as an electromagnetic modulator [41,51]. The second choice of temperatures is dictated by the steady state condition for the heat flux across the system, for which the net flux passing through slab 2 vanishes (see Section V).

## IV. MAGNETIC FIELD CONTROL OF HEAT FLUX ACROSS AN INTERMEDIATE MODULATOR

We are going to first consider the net heat flux received by slab 3 due exclusively to the thermal emission from slab 1, i.e., the role of the intermediate body only modulating the radiative fluxes passing through it, behaving as an electromagnetic signal modulator [41,51]. This situation can be achieved by initially setting the temperature of the whole system to *T*, while the temperature of body 1 is slightly increased to $T_1 = T+\Delta T$. Since now bodies 2 and 3 are still at the same temperature (i.e., $T_2 = T_3 = T$), there is no heat transfer between them, which, in turn, implies that $\mathcal{T}_{2\to 3}(\omega,\mathbf{k})$ in Eq. (9) does not play any role. Hence, Eq. (12) becomes

$$\varphi_3 = \int_0^\infty \frac{d\omega}{2\pi} \int \frac{d\mathbf{k}}{(2\pi)^2} \left[\Theta_1(\omega,T_1) - \Theta_3(\omega,T_3)\right] \mathcal{T}_{1\to 3}(\omega,\mathbf{k}). \qquad (13)$$

When $\Delta T \to 0$, we can introduce the linear conductance or the heat transfer coefficient *h*, defined as

$$h = \lim_{\Delta T \to 0} \frac{\varphi_3}{\Delta T} = \int_0^\infty \frac{d\omega}{2\pi} \int \frac{d\mathbf{k}}{(2\pi)^2} \frac{\partial \Theta(\omega,T)}{\partial T} \mathcal{T}_{1\to 3}(\omega,\mathbf{k}). \qquad (14)$$

Equation (14) becomes more compact than Eq. (13), and it only has one temperature *T*.

In Fig. 3 we summarize the numerical results of the heat transfer coefficient *h* as a function of the intermediate slab thickness *δ*, varying from 10 nm to 20 μm, for several values of the magnetic field intensity and of the gap size *d*. Moreover, to display the field effect more intuitively, we plot the ratio between the heat transfer coefficient with and without magnetic fields in the inset of each panel.

Let us first discuss the main features of the heat transfer coefficient *h* in the absence of a magnetic field (see black solid lines in all panels of Fig. 3). By comparing the results corresponding to different values of the gap size *d*, keeping the fixed thickness *δ*, one observes the usual decrease of the NFRHT as the vacuum gap overcome the length scale for the decaying of the SPP's and SPhP's evanescent fields, which dominate the RHT in the near-field regime. On the other hand, from these panels one can also observe the striking change in the dependence with the thickness δ of slab 2 of energy received by slab 3, as their separation increases. More specifically, in the cases of $d$ = 5 nm, 10 nm and 100 nm, there is a monotonically decreasing of the transfer coefficient $h$ as a function of δ [see Figs. 3(a)-3(c)]. This dependence is quite different from the observed one in Fig. 3(d), where $h$, corresponding to a 500 nm gap, exhibits a global maximum for a thickness around 700 nm. Such global maximum will be no longer present if $d$ is up to the far-field values (not shown here). This heat transfer behavior with respect to the intermediate slab thickness, is associated with the near-field effect occurring in three-body planar systems. In fact, as shown in Refs. [37,51], the intermediate slab may amplify the flux transferred from slab 1 to slab 3 when its thickness becomes comparable to the gap size (restricted to values of the near field regime), due to the optimal coupling of the SPP and SPhP cavity modes occurring in vacuum gaps.

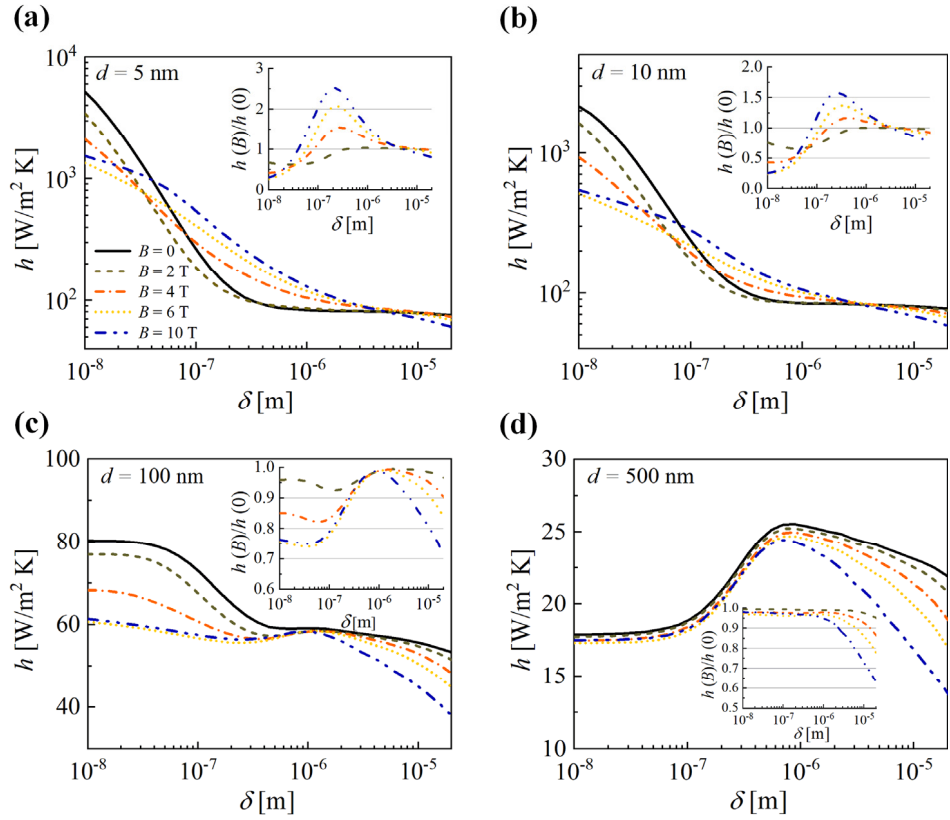

**Fig. 3.** Heat transfer coefficient $h$ as a function of the intermediate slab thickness δ for different values of the external magnetic field and of the gap size of (a) 5 nm, (b) 10 nm, (c) 10 nm and (d) 500 nm. The insets show the ratio between the heat transfer coefficients and their corresponding zero-field values. The temperature $T$ is set to 300 K.

As was mentioned in Sec. II, the NFRHT between two planar MO semiconductors is always reduced under the action of a magnetic field [23,31,32]. However, the situation becomes very different in the MO many-body configuration under study. For the small gap sizes, as $d$ = 5 nm and $d$ = 10 nm, the applied field has a capacity to reduce the heat transfer coefficient $h$ only when the intermediate slab thickness is close either to 10 nm or 20 μm, as evident from the insets of Figs. 3(a) and 3(b). When $d$ is increased to 100 nm, a field-induced reduction of the RHT is exhibited irrespective of slab 2 thickness, this reduction being more pronounced for both δ ≈ 10 nm and δ ≈ 20 μm [Fig. 3(c)]. As for higher values of $d$, the energy transfer modification is appreciable only for a thicker intermediate medium, as shown in Fig. 3(d).

On the other hand, we highlight that, for the small gap sizes, there is a region of values of δ, for which the presence of magnetic fields is able to enhance the NFRHT, showing that this peculiar amplification behavior occurs in the strong near-field regime. As presented in the inset of Fig. 3(a), corresponding to $d$ = 5 nm, such an enhancement can reach a value of 260% at $\delta \approx$ 215 nm, for a magnetic field intensity of 10 T. Incidentally, this enhancement value can be optimized by means of using a stronger external field (see Supplemental Material [59]), although high fields become challenging to achieve in practice [32,60]. This result marks a remarkable difference with the reported one for two-body planar InSb systems, implying an interplay between the HMs induced by magnetic fields and the cavity surface modes mediated by the intermediate medium.

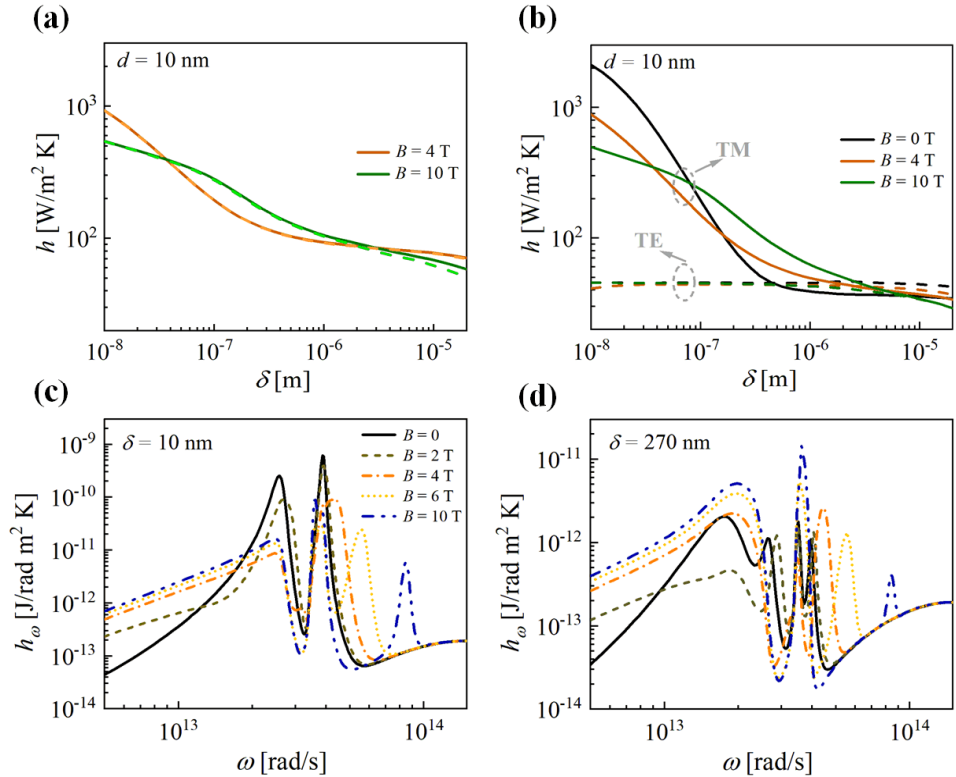

**Fig. 4.** (a) Heat transfer coefficient $h$ as a function of the intermediate slab thickness $\delta$, for magnetic field intensities of 4 T and 10 T. Solid lines correspond to the results from our method containing all mode contributions, while dashed lines correspond to the ones obtained neglecting polarization conversion. (b) Heat transfer coefficient $h$ for TE polarization (dashed lines) and TM polarization (solid lines) as a function of the thickness $\delta$ for magnetic field intensities of 0, 4 T and 10 T. (c) Magnetic field effects upon the TM spectral heat transfer coefficient $h_\omega$ for $\delta$ = 10 nm. (d) Same as in panel (c), but for $\delta$ = 270 nm. The gap size $d$ is 10 nm for all calculations.

To elucidate the physical mechanisms behind these modifications of the NFRHT induced by magnetic fields, it is necessary to analyze the contributions from all possible heat transfer channels, i.e., TE→TE, TE→TM, TM→TM and TM→TE. To this end, by using the analytic expression for isotropic materials presented in Ref. [37], we have calculated the heat transfer coefficient that neglects polarization conversion (i.e., TM→TE and TE→TM), and compare it with the results from our approach, accounting for all transmission channels. The results for two different strengths of magnetic fields are shown in Fig. 4(a), where we take the gap size $d$ = 10 nm. The good agreement between these results indicates that polarization conversion does not have a significative impact on the RHT. Meanwhile, it should be noted that, for high fields ($B = 10$ T) and large thickness ($\delta > 1$ μm), our approach allows us to obtain an exacter value of heat fluxes, owing to containing all transmission channels contributions. Subsequently, we further separate the TE and TM mode contributions in Fig. 4(b), where, as a reference, the zero-field result is added. From this plot it is manifest that the modifications to the RHT occurring for $\delta < 1$ μm can be traced to the magnetic field effect upon the transmission of TM-polarized evanescent waves. On the other hand, when the thickness of the intermediate slab becomes greater than 7.6 μm, we see that the flux contributed by TM polarization has decreased drastically, and becomes comparable to the one by TE polarization, arising from the fact that the tunneling of TM evanescent modes from slab 1 to slab 3 almost vanishes as a result of the extreme damping in the intermediate medium. This also renders the reduction of energy fluxes for TE modes be more appreciable, and comparable to that for TM ones.

Although we have shown that polarization conversion does not play a significant role, the underlying physics of the heat transfer modifications for TM modes, induced by the magnetic field in the thickness region $\delta < 1$ μm [see Figs. 3(b) and 4(b)], are still unexplained. For this, we now focus on the analysis of the magnetic field upon the TM spectral heat transfer coefficients for the thicknesses $\delta = 10$ nm and 270 nm, shown in Figs. 4(c) and 4(d), respectively. These results clearly indicate that the variation of energy fluxes caused by external magnetic fields is quite sensitive to the geometrical parameter of the modulator slab. Specifically, for a 10 nm intermediate slab, the applied field lessens the height of the zero-field peaks in the spectral flux associated with the SPP and SPhP modes, resembling the behavior for the two-body case [23,31], where the NFRHT is always reduced due to the substitution of SPP and SPhP modes by the field-induced HMs. In contrast, for a 270 nm slab, the field significantly increases the height as well as the width of the four zero-field peaks in the spectra, responsible for an effective enhancement of energy transferred to slab 3, as observed in the inset of Fig. 3(b).

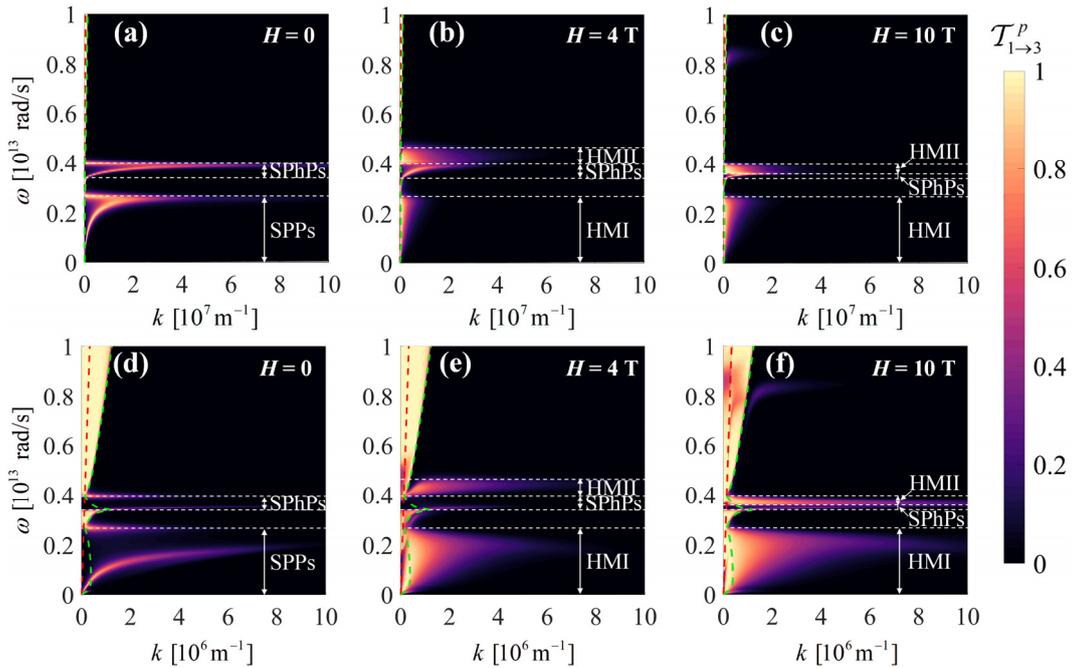

**Fig. 5.** Transmission probability for TM polarization from slab 1 to slab 3 as a function of the frequency $\omega$

and the parallel wave vector $k$ for the external magnetic field $B$ = 4 T and 10 T. Upper (lower) panels correspond to the intermediate slab thickness $\delta$ = 10 nm (270 nm). The gap size $d$ is 10 nm. Frequency regions corresponding to surface modes (SPPs and SPhPs) and hyperbolic modes are delimited by the horizontal white dashed lines. The red (green) dashed lines correspond to vacuum (InSb) the light line $\omega = ck$ ( $\omega = ck\sqrt{|\varepsilon_1|}$ ).

These features of the spectral flux can be understood by analyzing the TM-polarized transmission probability, $\mathcal{T}^p_{1\to3}$, shown in the upper (lower) panels of Fig. 5 for $\delta$ = 10 nm ($\delta$ = 270 nm), and for three different strengths of magnetic fields (i.e., $B$ = 0, 4 T, 10 T). Note that, in each panel, we plot the frequency regions that support SPPs, SPhPs, HMI and HMII modes when the InSb is exposed to the considered field strengths.

Let us first focus on the zero-field case. Comparing the zero-field transmission probability of Figs. 5(a) and 5(d) allows us to conclude that a very thin intermediate slab supports the formation of a cavity mode by coupling, inside of it, the evanescent fields from the surface modes at the interfaces of slabs 1 and 3. This can be well explained by analyzing the radiation penetration depth $l$. In the electrostatic limit, the penetration depth of an evanescent wave can be approximated by $l \approx k^{-1}$ [2]. Since only evanescent waves with $l \geq d$ can tunnel from slab 1 to slab 2, the largest contributing parallel wave vector $k_{max}$ between them, holding a major role in the near-field heat flux, can be approximated as $k_{max} \approx d^{-1}$ (a numerical proof is also provided in Supplemental Material [59]) with an associated penetration depth $l \approx d$. Therefore, for a 10 nm intermediate slab, we have $l \approx \delta$, favoring the coupling of surface modes at the two interfaces of the intermediate slab [61]. For this reason, the high transmission zones in Fig. 5(a) follow the dispersion law of cavity modes. In contrast, when the slab has a thickness of 270 nm (i.e., $\delta \gg l$), such a thick medium hinders the coupling of the evanescent modes coming from the outermost slabs, as demonstrated by the transmission spectra resembling uncoupled dispersion laws of single surfaces.

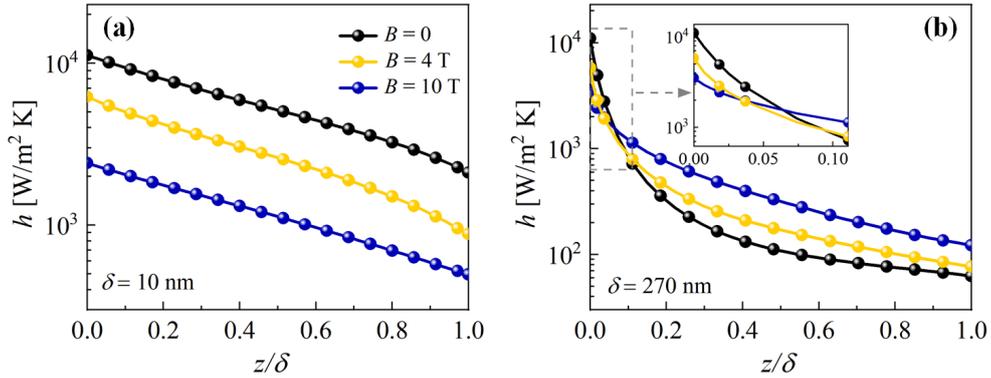

**Fig. 6.** Magnetic field dependence of the spatial profile of the TM-polarized energy flux across the intermediate slab with a thickness of (a) 10 nm and (b) 270 nm.

To display the influence of the intermediate slab thickness on the penetration depth of the evanescent modes more intuitively, in Figs. 6(a) and 6(b) we plot the spatial profiles of the absorbed heat (for TM polarization) inside the intermediate slab with $\delta$ = 10 nm and 270 nm, respectively. It can be clearly observed that, in the absence of a magnetic field (black lines), the energy flux decays exponentially inside the 270 nm slab as a result of the strong damping of the fields inside of it. This spatial dependence differs from the convex profile of the heat flux corresponding to $\delta$ = 10 nm, ascribed to the coupled evanescent fields coming from the outermost slabs.

We turn now to illustrate the variation of the transmission probability under external magetic fields. Figure 5 displays that, by increasing magnetic fields, the induced HMs gradually dominate energy transfer in the considered system, and they can benefit or not the photon tunneling, which depends on the intermediate slab thickness. Apparently, for $\delta = 10$ nm, the appearance of HMs will give rise to the destruction of the zero-field cavity mode between slabs 1 and 3. Besides, as shown in Figs. 5(a)–5(c), these field-induced HMs have significantly smaller wave vector cutoffs relative to the disappeared surface modes, responsible for the reduced height of spectral peaks under the action of magnetic fields in Fig. 4(a). Therefore, this mechanism for the field-induced-reduction of the NFRHT resembles the reported one for two-body InSb configurations [23,31]. This can be further supported by observations of Fig. 6(a). First of all, we see that, in fact, the presence of magnetic fields has strongly suppressed the energy flux produced by tunneling into slab 2 (corresponding to position $z/\delta = 0$), especially high fields, which is exactly due to the generation of the HMs that greatly takes the place of those zero-field surface modes with significantly larger wave vectors. Secondly, for such a small thickness, the evanescent fields are weakly damped in the intermediate body (as discussed before). These two factors jointly determine that, within the intermediate slab, the energy flux at finite fields is always smaller with respect to the zero-field case [Fig. 6(a)]. Thus, one finally sees that the magnetic field reduces the transmission probability towards slab 3 due to the appearance of HMs. However, when the slab thickness $\delta$ is 270 nm, we first see in Fig. 6(b) that, in contrast to the zero field, the field leads to a significantly slower drop of the radiative flux inside the intermediate slab, although still suppressing the flux entering into the slab strongly. Let us remind that the evanescent fields coming from the surface modes at the interfaces of slabs 1 and 3 do not couple inside such a thick intermediate slab to form the zero-field cavity mode. Therefore, the role of the magnetic field is now to covert a low flux tunneling, arising from the strong damping of the uncoupled surface modes in the intermediate medium, into an efficient tunneling mediated by the hyperbolic modes, whose capability to propagate increases with the applied field intensity, as evidenced by the profiles of Fig. 6(b). This, in turn, makes those field-induced HMs exhibit broader bands and even larger wave vector cutoffs relative to the zero-field surface modes they replace, as observed in Figs. 5(d)–5(f).

## V. INTERMEDIATE MEDIUM PRODUCING ADDITIONAL CONTRIBUTIONS IN STEADY STATE

In the previous section, we have demonstrated that an applied magnetic field can not only reduce but also enhance the RHT in the MO three-body planar configuration, associated with the interplay between the damped evanescent fields of the surface waves and the propagating hyperbolic modes induced by external magnetic fields. Notably, this finding is achieved under the assumption of thermal equilibrium between slabs 2 and 3. Therefore, it is natural to ask ourselves if such a field-induced enhancement, obtained for $d = 10$ nm and $\delta = 270$ nm, persists when the intermediate medium starts providing an additional contribution (i.e, $T_2 \neq T_3$, so that $\mathcal{T}_{2\to3}(\omega,\mathbf{k})$ plays a role). Here, from a practical point of view, we are going to consider a general steady-state situation [37,49,53]. Specifically, we assume that the external slabs 1 and 3 are still held at temperatures $T_1$ and $T_3$, respectively, while the intermediate slab now relaxes into steady state, and thus has an equilibrium temperature $T_2$. This temperature can be solved by using the thermal equilibrium condition that the net energy fluxes to and from slab 2 should be zero, i.e.,

$$\Delta\varphi = \varphi_{1\to 2} - \varphi_{2\to 3} = 0. \tag{15}$$

Here, $\varphi_{1\to 2} = \int_0^\infty \frac{d\omega}{2\pi} \int \frac{d\mathbf{k}}{(2\pi)^2} [\Theta_1(\omega,T_1) - \Theta_2(\omega,T_2)] \mathcal{T}_{1\to 2}(\omega,\mathbf{k})$ represents the net flux from slab 1 to slab 2, while $\varphi_{2\to 3} = \int_0^\infty \frac{d\omega}{2\pi} \int \frac{d\mathbf{k}}{(2\pi)^2} [\Theta_2(\omega,T_2) - \Theta_3(\omega,T_3)] \mathcal{T}_{2\to 3}(\omega,\mathbf{k})$ represents that from slab 2 to slab 3. Note that, since the system under consideration is symmetric with respect to slab 2, we always have $\mathcal{T}_{1\to 2}(\omega,\mathbf{k}) = \mathcal{T}_{2\to 3}(\omega,\mathbf{k})$, regardless of the presence of magnetic fields. Based on this, Eq. (15) can be rewritten as

$$\Delta\varphi = \int_0^\infty \frac{d\omega}{2\pi} \int \frac{d\mathbf{k}}{(2\pi)^2} [\Theta_1(\omega,T_1) + \Theta_3(\omega,T_3) - 2\Theta_2(\omega,T_2)] \mathcal{T}_{2\to 3}(\omega,\mathbf{k}) = 0. \tag{16}$$

From Eq. (16) it is clear that, once one determines the energy transmission coefficient $\mathcal{T}_{2\to 3}(\omega,\mathbf{k})$ as well as the temperatures $T_1$ and $T_3$, the equilibrium temperature $T_2$ can be easily solved. Note that the value of $\mathcal{T}_{2\to 3}(\omega,\mathbf{k})$ will be changed for different strengths of the applied field. In the following we take the temperatures $T_1 = 400$ K and $T_3 = 300$ K, and assume that the optical properties of the InSb remain unchanged for 400 K.

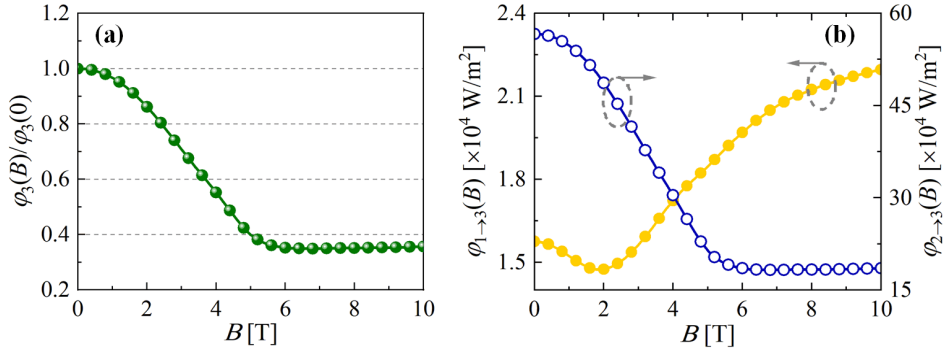

**Fig. 7.** (a) Total steady-state heat flux absorbed by slab 3 as a function of the magnetic field and normalized to the zero-field value. (b) Net heat fluxes $\varphi_{1\to 3}$ and $\varphi_{2\to 3}$ transferred from slabs 1 and 2 to slab 3, as a function of the magnetic field intensity. The temperatures of slab 1 and slab 3 are set to $T_1 = 400$ K and $T_3 = 300$ K, respectively. The gap size $d$ is 10 nm, and the intermediate slab has a thickness of $\delta = 270$ nm.

In Fig. 7(a), we show the total steady-state flux received by slab 3 (given by Eq. (12)) as a function of the magnetic field intensity and normalized to its zero-field value. We clearly see a monotonically decrease in the energy flux with the applied field, until it saturates for fields above 6 T. One would think that this reduction of the RHT is due entirely to the full replacement of SPP and SPhP modes by HMs, as occurs for two semi-infinite InSb plates [23]. However, this is not completely true, as exhibited in Fig. 7(b), where it can be observed that the net heat fluxes from slabs 1 and 2 to slab 3, exhibit different dependences on the field intensity. Specifically, the flux $\varphi_{1\to 3}$ has a growing dependence with the field, consistent with the enhancement effect mediated by the intermediate slab, as discussed in the previous section. On the other hand, the flux $\varphi_{2\to 3}$ involves the direct exchange of heat between slabs 2 and 3, for which it is not surprising to observe a decreasing dependence with the applied field, similar to the two-body case [23]. Nevertheless, owing to the proximity of slabs 2 and 3, the flux transferred between them overcome by almost an order of magnitude that between slabs 1 and 3. Consequently, as the field intensity increases, the reduction of $\varphi_{2\to3}$ always dominates the increase of $\varphi_{1\to3}$, resulting in the net decreasing flux arriving at slab 3.

## VI. CONCLUSIONS

In conclusion, we have theoretically investigated the magnetic field dependence of the near-field heat transfer in a three-body system made of MO *n*-doped InSb slabs. A general Green-function-based approach for the analysis of the RHT in many-body planar systems made of arbitrary materials has been provided. We employed it to exactly calculate the radiative heat flux produced in the considered MO three-slab configuration, and displayed that, when the intermediate body acts as an electromagnetic modulator in thermal equilibrium with one of the outermost slabs, the presence of the magnetic field is capable of giving rise to either a reduction or an enhancement of the radiative energy transfer. By analyzing the photon transmission probability as well as the spatial profiles of the absorbed heat inside the intermediate slab, we demonstrated that this peculiar NFRHT enhancement occurs when the thickness of this slab is large enough to hinder the coupling of the evanescent fields entering on, leading to a low photon tunneling, which, in turn, can be enhanced by the propagating HMs induced by the magnetic field. In addition, we pointed out that this field-induced enhancement effect on the energy transfer will disappear if the whole system reaches steady state, since a net decrease of the heat flux arriving at the cold slab is achieved due to the additional flux it receives from the intermediate medium.

Our method and results pave the way to further investigating some interesting thermomagnetic effects in systems with a large number of planar slabs of MO materials. For example, in such a complex system, the giant thermal magnetoresistance effect are still unknown. In this regard, an important question is whether this effect is achievable with a very low magnetic field, which could be crucial for efficiently controlling energy transfer in practical applications. Moreover, the temporal dynamics of radiative heat transfer in complex many-body systems has a growth of interest recently [62–64]. In this respect, it may be very interesting to investigate how the external magnetic field affects the thermalization dynamics for a system consisting of many MO slabs, which is instructive for developing ultrafast thermal management devices.

**Acknowledgements**

We thank I. Latella for fruitful discussions. This work was supported by the National Natural Science Foundation of China (Grant No. 52076056) and by the Fundamental Research Funds for the Central Universities (Grant No. FRFCU5710094020).